\documentclass[sigconf]{acmart}

\usepackage{booktabs} 
\usepackage{graphicx}
\usepackage{subfigure}
\usepackage{CJKutf8} 
\setcopyright{rightsretained}

\usepackage{hyperref}
\usepackage{color}
\hypersetup{colorlinks,
  urlcolor=[rgb]{0.5,0.5,0.5},
            citecolor=blue,
            linkcolor=blue}
\acmISBN{} 
\def\acmDOI#1{\def\@acmDOI{#1}}
\acmDOI{} 

\acmConference[LBRS@RecSys'18]{the Late-Breaking Results track part of the Twelfth ACM Conference on Recommender Systems (RecSys’18) }{October 2--7, 2018}{Vancouver,BC,Canada}

\begin{document}
\title{Representation Learning for Image-based Music Recommendation}


\author{Chih-Chun Hsia}
\affiliation{%
 \institution{National Chengchi University}
 \city{Taipei} 
 \state{Taiwan} 
}
\email{g10401@cs.nccu.edu.tw}

\author{Kwei-Herng Lai}
\affiliation{%
 \institution{Academia Sinica}
 \city{Taipei} 
 \state{Taiwan} 
}
\email{khlai@citi.sinica.edu.tw}

\author{Yian Chen}
\affiliation{%
 \institution{KKBOX Inc.}
 \city{Taipei} 
 \state{Taiwan} 
}
\email{annchen@kkbox.com}

\author{Chuan-Ju Wang}
\affiliation{
  \institution{Academia Sinica}
  \city{Taipei}
  \state{Taiwan}
}
\email{cjwang@citi.sinica.edu.tw}

\author{Ming-Feng Tsai}
\affiliation{
  \institution{National ChengChi University}
  \city{Taipei}
  \state{Taiwan}
}
\email{mftsai@nccu.edu.tw}

\renewcommand{\shortauthors}{C.Hsia, K. Lai, Y. Chen, C. Wang, and M. Tsai}

\begin{abstract}
Image perception is one of the most direct ways to provide 
contextual information about a user concerning his/her surrounding environment; 
hence images are a suitable proxy for contextual recommendation.
We propose a novel representation learning framework for image-based music
recommendation that bridges the heterogeneity gap between music and image data;
the proposed method is a key component for various contextual recommendation
tasks.
Preliminary experiments show that for an image-to-song retrieval task,
the proposed method retrieves relevant or conceptually
similar songs for input images.

\end{abstract}
\vspace{-0.2cm}

\vspace{-0.4cm}
\vspace{-0.4cm}

\maketitle

\vspace{-0.2cm}
\section{Introduction}
\vspace{-0.1cm}
%
%

Contextualization is an important feature in many recommender
systems~\cite{jannach2016recommender}.
For example, music-streaming services such as Spotify, use the
current mood of the user or adapt recommendations depending on the time of the
day. Apart from the user mood or time information, image perception is a more
direct way to acquire contextual information about the user's immediate
environment due to its rich information. Therefore, images are a better proxy 
for contextual recommendation. 
However, the heterogeneity gap between different types of data makes this challenging.
In this paper we propose an innovative framework to bridge the heterogeneity gap 
between music and image data for image-based music recommendation; as such, the proposed 
method is a key component for various contextual recommendation tasks.

Figure~\ref{fig:framework} shows the three modules of the proposed framework:
1) a CNN module, 2) a network embedding module, and 3) a retrieval module.
The CNN module uses the VGG-19~\cite{VGG-19} pre-trained model to obtain the
image representations.
The network embedding module, inspired
by~\cite{Wang:2017:IIC:3077136.3080807}, bridges the concepts of songs
and images via keywords, and learns the representations of images, songs, and
their corresponding keywords based on neighborhood
proximity~\cite{LINE,Wang:2017:IIC:3077136.3080807}.
Finally, the retrieval module yields the recommended songs for an input image
given the representations from the CNN and network embedding modules.
The experiments show that for the image-to-song retrieval task, given the input image,
 the proposed method retrieves songs that are relevant or conceptually similar.

\vspace{-0.2cm}
\section{Image-based Music Recommendation}
\vspace{-0.1cm}

\begin{figure*}[ht]
  \centering
  \graphicspath{{source/}}
  \includegraphics[height=5cm]{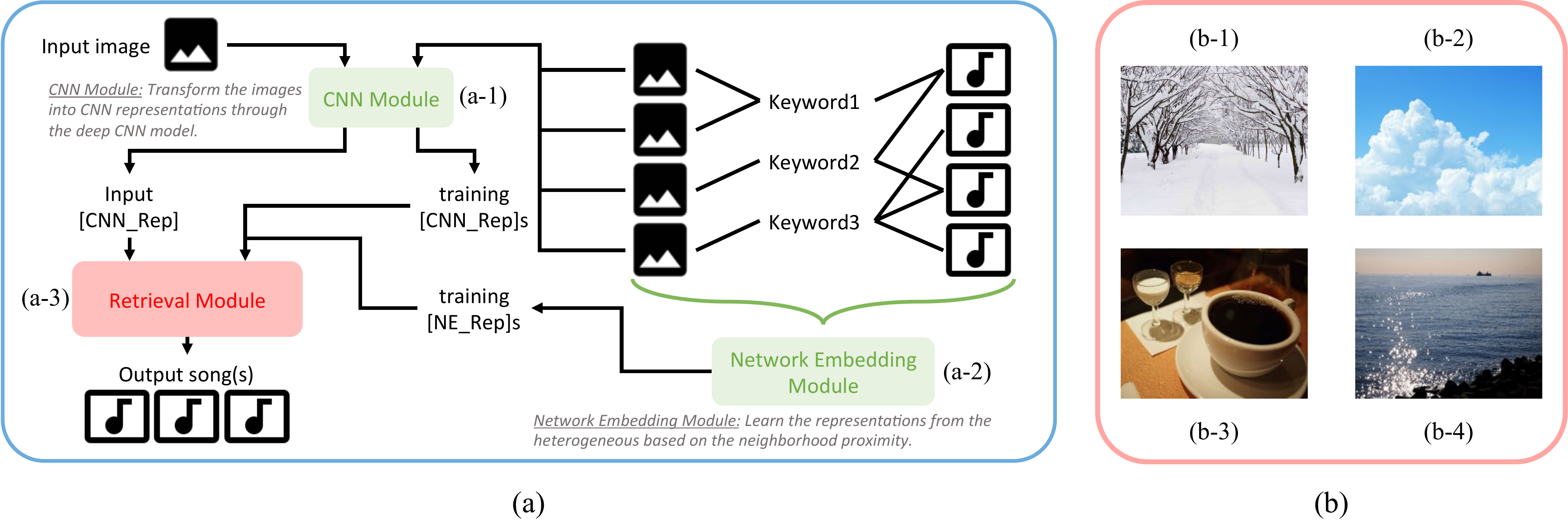}
  \caption{Proposed framework and example cases}
  \label{fig:framework}
  \vspace{-0.6cm}
\end{figure*}

Figure~\ref{fig:framework}(a) illustrates the three modules of the proposed
framework.


\emph{CNN Module}
In this framework we apply the VGG-19 pre-trained model, which yielded a 7\%
top-5 error on the ILSVRC-2012 dataset, to obtain the image representations
(hereafter referred to as the CNN-based representations).
The network structure of VGG-19 includes 16 convolutional layers and 3
fully-connected layers, with the use of $3\times3$ filters.
To generate the 4096-dimensional representation for each image,
we extract the representations from the second rather than the third fully-connected layer.
In our task, we use weights pre-trained on a~1000-class object recognition task
using about 150,000~$224\times224$-pixel images.

\emph{Network Embedding (NE) Module}
We use neighborhood proximity~\cite{LINE,Wang:2017:IIC:3077136.3080807} 
in network embedding to capture the relationship between images and songs, 
and then use this relationship for recommendations.
In particular, we construct a heterogeneous tripartite network by connecting
the two types of multimedia data with corresponding keywords; hence there are
three types of vertices (i.e., images, words, and songs) and two types of edges
in the network:
\vspace{-0.2cm}
\begin{enumerate}
  \item Song-keyword edge: connects each song with the keywords in its lyrics; 
    the weight indicates the relevance between the
    song and the keyword.
  \item Image-keyword edge: connects each image with its corresponding keyword; 
    the weight indicates the relevance between the image and the keyword.
\end{enumerate}
\vspace{-0.2cm}
In the proposed framework, the vertex representations are learned based on
their neighborhood proximity using stochastic gradient descent with edge
sampling~\cite{LINE} and negative sampling~\cite{mikolov2013distributed}.

\emph{Retrieval Module}
Given an input image, the retrieval module comprises the following three
stages:
\vspace{-0.1cm}
\begin{enumerate}
  \item Image transformation: the CNN module is used to transform the input image into a 4096-dimensional
    CNN-based representation.
  \item Image retrieval: the most relevant images with respect to the input image 
    are retrieved by calculating the Euclidean distance between their CNN-based representations and those
    of the pre-trained images.
  \item Song recommendation: the most relevant songs are recommended based on
    the Euclidean distances between the NE-based representations of songs and
    the images retrieved in the previous stage.
\end{enumerate}
\vspace{-0.1cm}

\vspace{-0.1cm}
\section{Preliminary Experiments}
\vspace{-0.1cm}
We obtained the music dataset from KKBOX\footnote{\url{http://www.kkbox.com}};
and crawled the lyrics for keyword extraction and network construction for the
proposed framework.
For the keyword extraction, we used the Jieba toolkit\footnote{\url{https://github.com/fxsjy/jieba}} for Chinese word
segmentation and extracted 72 frequent keywords from the titles and song lyrics.
Moreover, we constructed our image dataset by using search engines to collect images 
given the 72 selected keywords, and also included in the experimental dataset songs that
contained at least one of the keywords, yielding a total of 62,316 songs, 72 keywords, and 33,459 images.

\vspace*{-0.4cm}
\begin{table}[tbh]
	\caption{Image-to-song retrieval (hit rate@10)}
	\vspace*{-0.4cm}
	\centering
	\setlength{\tabcolsep}{10pt}
	\label{tb:het_exp}
	\begin{tabular}{cccccc}
		\toprule
                        $n$    & Proposed model & KM & POP\\
                \midrule                                  
                        5      & \bf{0.913}     & 0.902           & 0.124  \\ 
                        10     & \bf{0.918}     & 0.917           & 0.157  \\
                        50     & \bf{0.943}     & 0.941           & 0.356  \\
                        100    & \bf{0.943}     & 0.943           & 0.378  \\
		\bottomrule
	\end{tabular}
\end{table}
\vspace*{-0.4cm}

Table~\ref{tb:het_exp} lists the results of an image-to-song retrieval task.
For each input image, our framework recommended the top 10 songs obtained from
top 2 relevant songs for the top 5 relevant images, whereas the keyword matching
(KM) baseline recommended 10 songs directly based on the keyword associated with 
the input image in songs' lyrics and the popular (POP) baseline recommended 10 songs 
randomly selected from the top-100 most popular songs.
For each of the 72 keywords, we collected the top $n$ conceptually similar
words of the extracted keywords from
ConceptNet\footnote{\url{http://conceptnet.io}} to construct the ground truth:
the lyrics of a recommended song containing at least one of the associated
keywords or their corresponding conceptually similar words.
The result shows that our framework is able to retrieve more
literally relevant songs to the input images than the two baselines, KM and
POP.


Furthermore, we conducted an user study in which we collected feedback from 10
 users, by having each user listen to the top 10 retrieved songs for each of 
the given~4 input images.
Figure~\ref{fig:framework}(b) shows the 4 input images and Table~\ref{tb:user}
tabulates the results in terms of precision; the performance of our method is superior
to all of the baseline methods.  
For snow-themed images, it is worth mentioning that the lyrics of
most of the recommended songs included the concept-related words
\emph{Christmas} and \emph{cold}; this shows that the proposed method
recommends music based on contextual information.
To sum up, Tables~\ref{tb:het_exp} and ~\ref{tb:user} indicate that our method
is able to retrieve more relevant songs than the baseline methods in terms of
both literal and contextual aspects.
\vspace{-0.3cm}
\begin{table}[h]
  \caption{User feedback for the 4 images (precision@10)}
	\vspace*{-0.4cm}
	\label{tb:user}
	\setlength{\tabcolsep}{3pt}
	\begin{tabular}{ccccc}
		\toprule
		 Theme & Snow forest & Sky with clouds & Coffee & Ocean\\
		   & (b-1) & (b-2) & (b-3) & (b-4) \\ 
		\midrule
                 Proposed model & \bf{0.776} & \bf{0.623} & \bf{0.655} & \bf{0.709}\\
                 KM             & 0.531 & 0.569 & 0.546 & 0.531     \\
		 POP            & 0.414 & 0.514 & 0.371 & 0.586	\\
		\bottomrule
	\end{tabular}
\end{table}
\vspace*{-0.6cm}

\section{Conclusion and Future Work}
\vspace{-0.1cm}
We present a representation learning framework that bridges the
heterogeneity gap between music and image information for image-based music
recommendation.
Preliminary results demonstrate that the proposed method retrieves more
relevant or conceptually similar songs for input images than the baseline
methods.
In future, we will incorporate user listening behavior into the
NE module to construct an image-based and personalized recommender system.

  



\vspace{-0.3cm}
\bibliographystyle{ACM-Reference-Format}
\bibliography{paper} 

\end{document}